# Error analysis and lattice improvement for the C-ADS Injector-I*

MENG Cai(孟才), LI Zhi-Hui(李智慧), TANG Jing-Yu(唐靖宇)
Institute of High Energy Physics, Chinese Academy of Sciences, Beijing 100049, China

**Abstract** The injector Scheme-I (or Injector-I) of the C-ADS linac is a 10-mA 10-MeV proton linac working in CW mode. It is mainly comprised of a 3.2-MeV room-temperature 4-vane RFQ and twelve superconducting single-spoke cavities housed in a long cryostat. Error analysis including alignment and field errors, static and dynamic ones for the injector are presented. Based on detailed numerical simulations, an orbit correction scheme has been designed. It shows that with correction the rms residual orbit errors can be controlled within 0.3 mm and a beam loss rate of $1.7\times10^{-6}$ is obtained. To reduce the beam loss rate further, an improved lattice design for the superconducting spoke cavity section has been studied.

## 1 Introduction

The C-ADS (China Accelerator-Driven Subcritical System) project is a strategic plan to solve the nuclear waste and resource problems for nuclear energy in China [1]. The C-ADS accelerator is a CW proton linac and uses superconducting acceleration structures except the RFQs and consists of two injectors and a main linac section, as shown in Fig. 1.

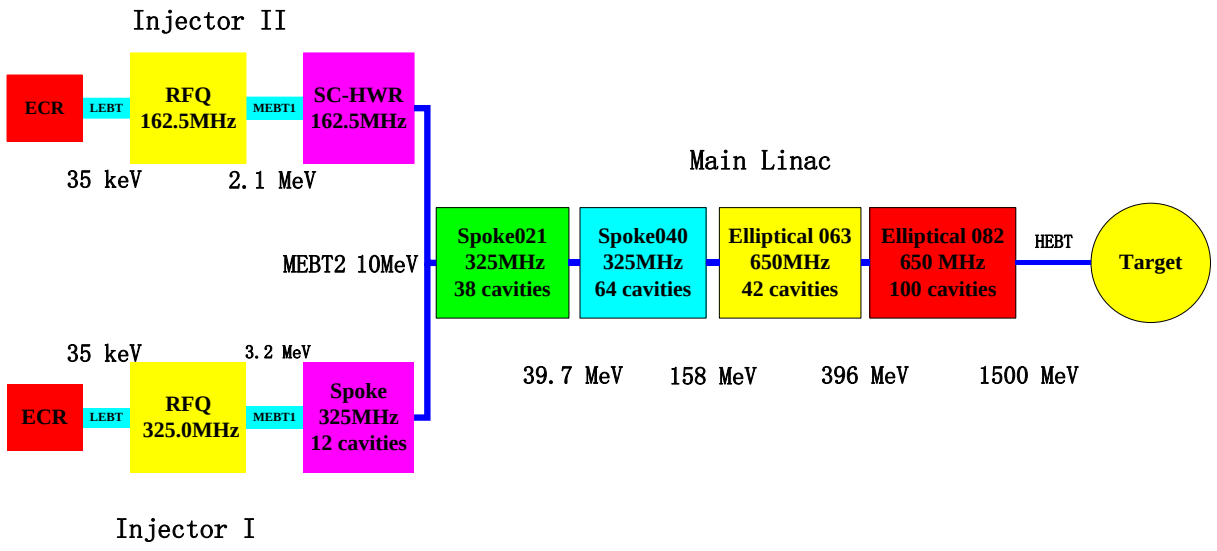

Fig. 1 Layout of the C-ADS driver accelerator.

Two identical injectors will be operated in the mode of one as the hot-spare of the other. However, two different injector schemes are shown in Fig.1, and this means that in the early developing phase two different approaches of injector will be developed in parallel by two teams. The Injector Scheme-I [1] uses a 3.2-MeV normal conducting 4-vane RFQ and twelve superconducting single-spoke cavities (Spoke012 type), as shown in Fig. 2. It is divided into four sections: ion source – LEBT (low-energy beam transport) section, RFQ section, MEBT1 (medium-energy beam transport) section and Spoke012 section. This paper will represent the studies about the error analysis and orbit correction scheme in the injector including MEBT1 and the spoke cavity section. Based on the error studies, the improved lattice design for the superconducting section to reduce beam loss rate is also presented, which has a larger tolerance on errors.

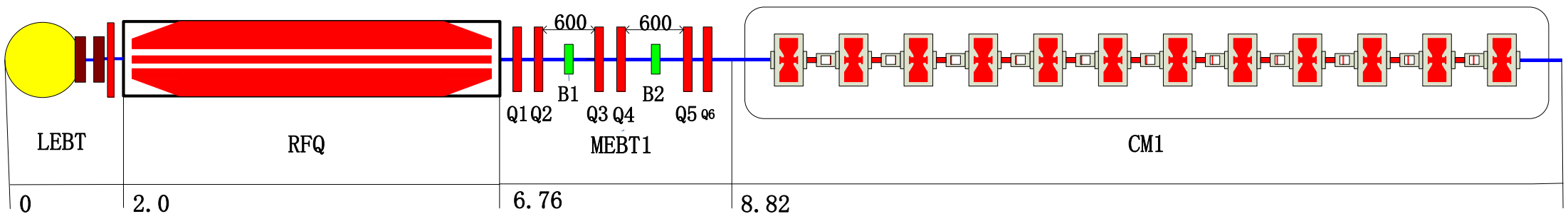

Fig. 2. Schematic of the Injector Scheme-I lattice.

* Supported by the China ADS Project (XDA03020000)
E-mail: tangjy@ihep.ac.cn, mengc@ihep.ac.cn

## 2 Simulations with errors

### 2.1 Error sources

All the devices having electromagnetic field influence over the beam should have installation errors including translational errors and rotational errors, and also field errors. We can classify the possible error sources into three groups [2]:

1) Misalignment errors: affecting all the elements with translational errors and rotational errors, e.g. solenoids, quadrupoles, accelerating cavities, etc.
2) Field errors: affecting the field levels as well as the phases of RF accelerating cavities and the fields of magnets.
3) BPM uncertainty errors: affecting the orbit correction effect.

All the errors mentioned above can be also classified in two different types according to their variation properties with time: static errors and dynamic errors. For the injector, misalignment errors are basically the static ones, as the influence of the dynamic errors on the residual orbit errors are small. The field errors of magnets and RF cavities should be considered as both static errors and dynamic errors. In a real machine, the effect of static errors can be partially corrected with the help of beam measurements. In the simulations presented here, we do not distinguish between static and dynamic errors of RF fields.

### 2.2 Error settings

Following the engineering experience and the special requirements for a superconducting linac, the errors used for error studies are shown in Table 1. The errors are generated randomly between minus and positive value in Table 1with uniform distribution, which is similar as in other linac studies [3-5]. The uniform error distributions with the amplitudes shown in the table give relatively more pessimistic results than truncated Gaussian distributions. In the simulations, 1000 sets of errors are generated and applied to the corresponding elements and $10^5$ particles are tracked for each set, which makes a total of $10^8$ particles for one simulation for error analysis. Although the beam loss rate is the most critical factor in error analysis, other beam parameters such as residual orbit errors and emittance growth, are used to analyze the influence of different types of errors and the effectiveness of the orbit correction scheme.

Table 1. Amplitudes of errors used for error studies.

| Error No. | Error description | Tolerance Static | Dynamic |
|---|---|---|---|
| 1 | Magnetic element displacement | | |
| | Quadrupole | 0.1 mm | 2 μm |
| | Solenoid (cold) | 1 mm | 10 μm |
| 2 | Magnetic element rotation | 2 mrad | 0.02 mrad |
| 3 | Magnetic element field | 0.5 % | 0.05% |
| 4 | Cavity displacement (cold) | 1 mm | 10 μm |
| 5 | Cavity rotation | 2 mrad | 0.02 mrad |
| 6 | RF amplitude fluctuation | 1% | 0.5% |
| 7 | RF phase fluctuation | 1° | 0.5° |
| 8 | BPM uncertainty | 0.1mm | |

### 2.3 Error sensitivity analysis

To study the sensitivities of errors, we set bigger errors than those listed in Table.1. The effects of errors on the residual orbit errors at the injector exit without correction scheme are shown in Fig.3. The effects of different errors relative to the inputs are plotted as functions of error tolerances. The residual orbit error is expressed as $\sqrt{\frac{1}{N}\sum_{i}^{N} A_i}$ , where $A_i$ is residual orbit error with the i$^{th}$ run and $N$ is the number of runs. We can find that the transverse residual orbit errors are mainly affected by misalignment errors of solenoids and cavities. The energy jitter is mainly affected by RF field errors and solenoid displacements. The integral of absolute value of the electric field (Ez) for a superconducting spoke cavity (Spoke012) is shown in Fig.4. One can find that the field variation with respect the transverse position is quite large.

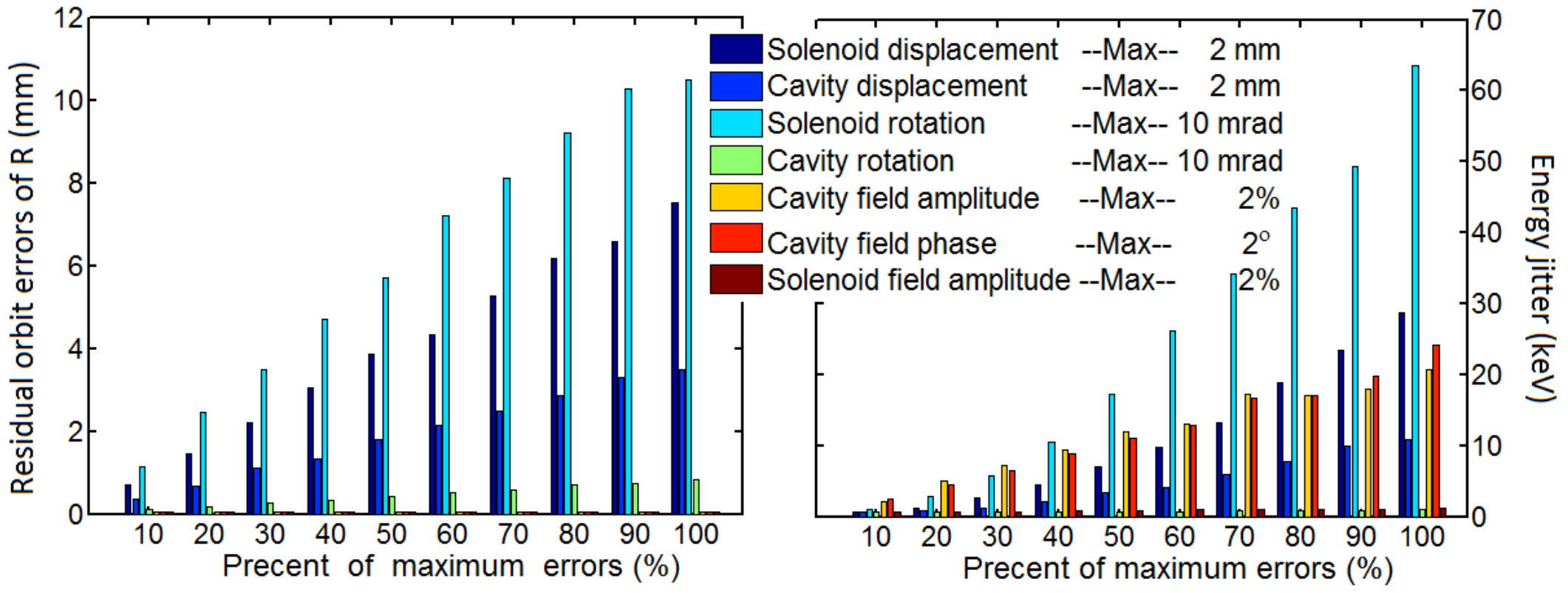

Fig. 3. (color online) The residual orbit errors with different errors.

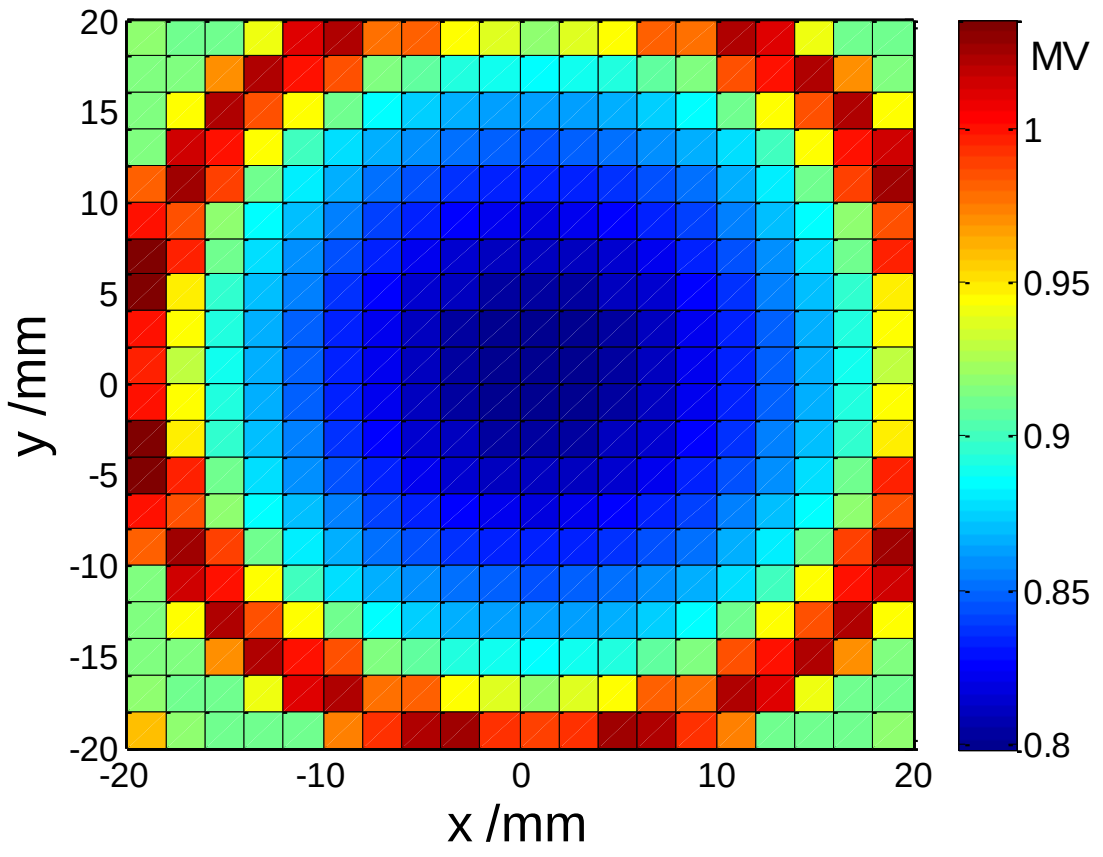

Fig. 4. (color online) Integral of the electric field |Ez| in a Spoke012 cavity

The emittance growth and energy jitter with the orbit correction are shown in Fig. 5. In this paper the emittance refers to rms emittance. We can find that the transverse emittances are mainly affected by cavity displacements (Error 4) and magnetic field ripples (Error 3), and the longitudinal emittance is mainly affected by cavity displacements (Error 4) and RF field errors (Errors 6 and 7). From Fig.4 and above results we can see that the field variations affect the beam quality greatly. So the cavity displacements are very important errors in the injector, the procedure of cavity installation should be carefully designed.

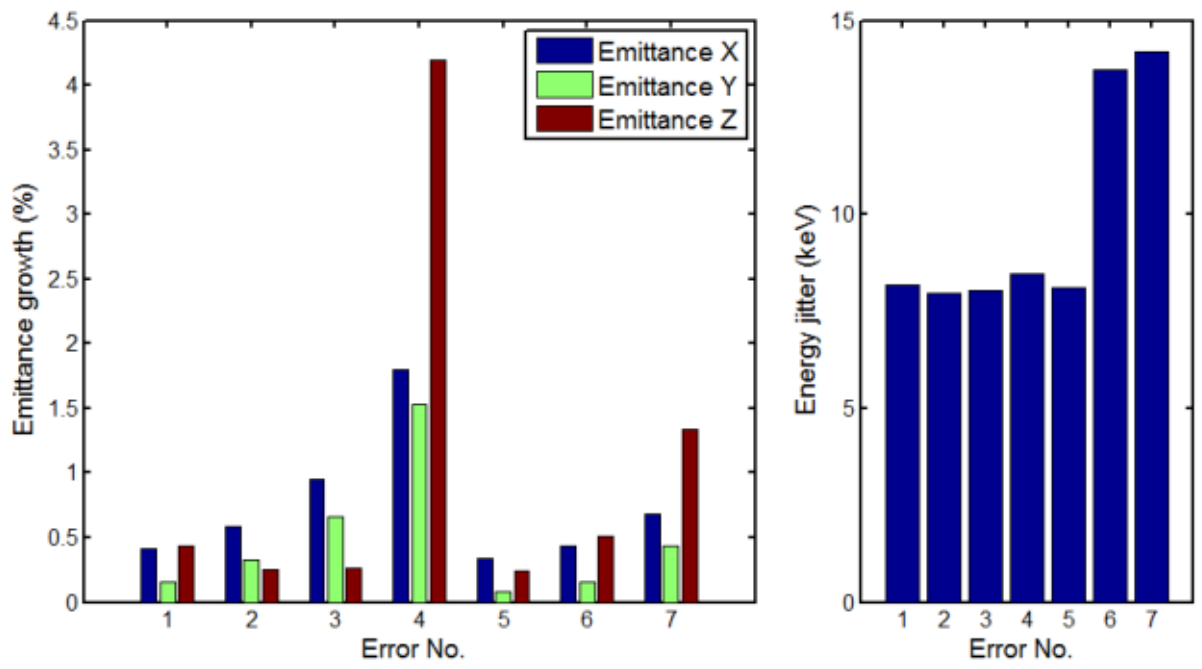

Fig. 5. (color online) Emittance growth and energy jitter with different errors in Table 1.

## 2.4 Orbit correction scheme for the Spoke012 section

The multi-particle simulations show that the residual orbit errors are too large without correction, and it will result in evident beam loss and beam quality degrading. Thus a good orbit correction is required. According to the lattice design, the transverse phase advance per period is 40-70°, a pair of corrector and BPM (beam position monitor) in each period is arranged for the orbit correction. The correction scheme in the MEBT1 section uses the coils attached to the quadrupoles and the BPMs which were presented in Ref. [7]. The correction scheme in the Spoke012 section relies on the steering coils attached to the solenoids and the BPMs, as shown in Fig. 5. This one-to-one correction scheme maintains the RMS residual orbit errors within 0.4 mm while keeping the maximum deviation within 1 mm and the RMS emittance growth below 10%, as shown in Fig. 6. The BPMs can be aligned up to a few tens micrometres by BBA method [6] and the reading noise of BPMs is about 30 μm, but we adopted the BPM uncertainty of 0.1 mm in the simulations.

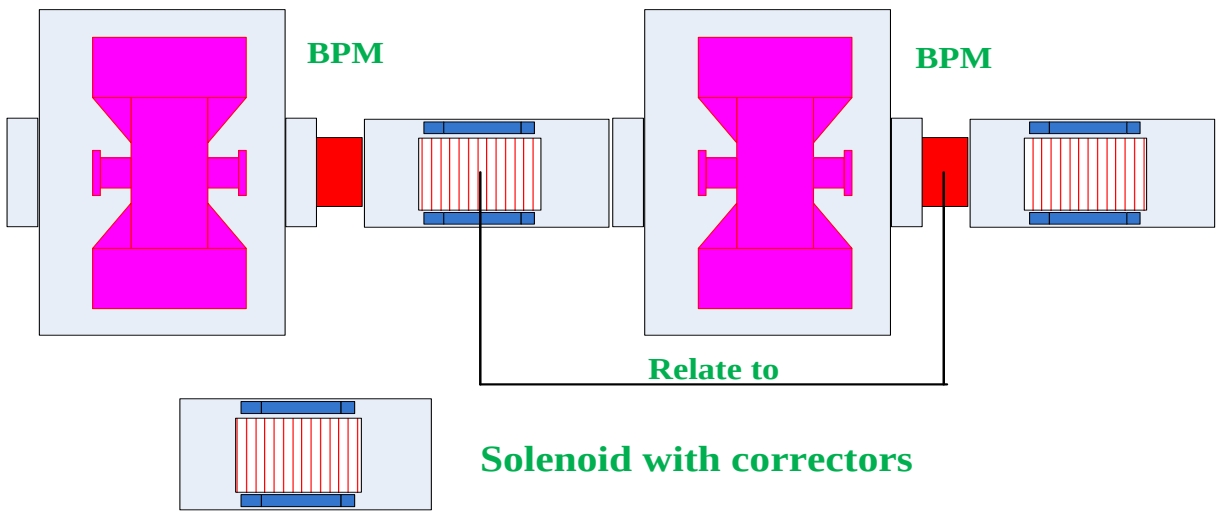

Fig. 6 Schematic of the orbit correction scheme for the Spoke012 section in the injector.

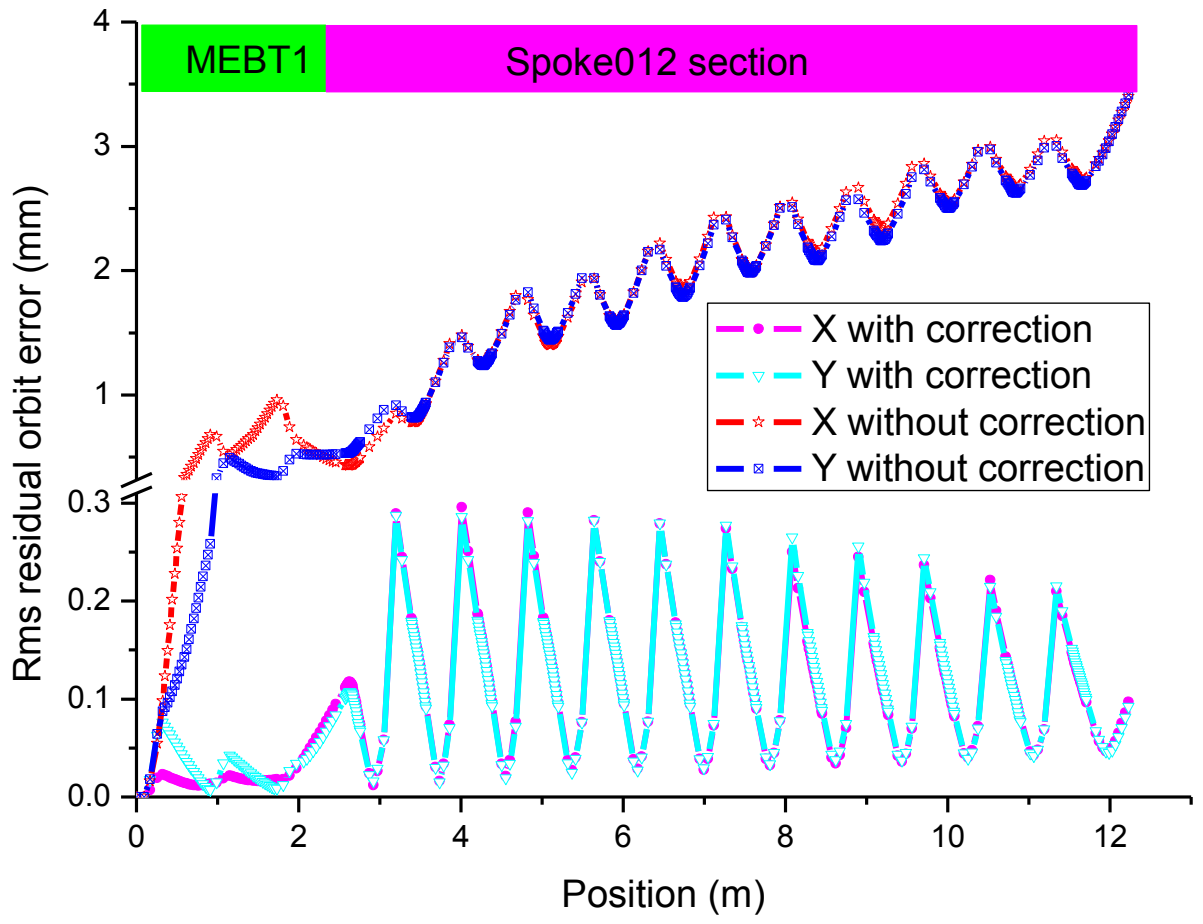

Fig. 7. (color online) Simulation results with nominal errors in the injector.

## 2.5 Other simulation results

The tracking of $10^5$ particles using the simulated RFQ exit distribution was carried out through 1000 different linacs with errors and orbit corrections. It turns out that the orbit correction scheme works well by controlling the rms residual orbit error within 0.4 mm. The beam loss rate is about $1.7\times10^{-6}$. The particle trajectories in the horizontal and longitudinal planes along the MEBT1 and the SC section are shown in Fig. 8. We can see that some particles move out of the longitudinal acceptance in the SC section, and then they will not match to the transverse focusing channel downstream and get lost finally. The relatively small longitudinal acceptance of the Spoke012 section is considered the main reason for the beam loss. In order to verify this, we studied the relation between the beam loss rate and the RF errors which is summarized in Table 2. Beam losses with different initial distributions with errors are shown in Fig. 9. Here, the initial beam distribution is a truncated Gaussian distribution. We can see that the initial longitudinal distribution has a great influence over beam loss, which also indicates that the longitudinal acceptance of the Spoke012 section is relatively too small.

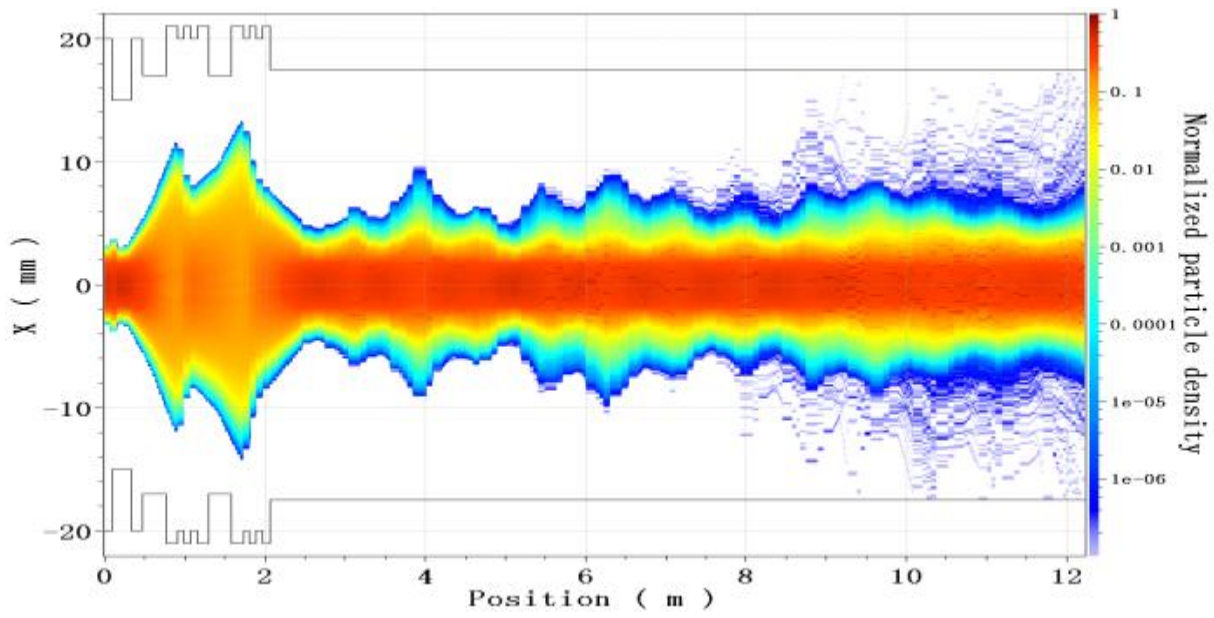

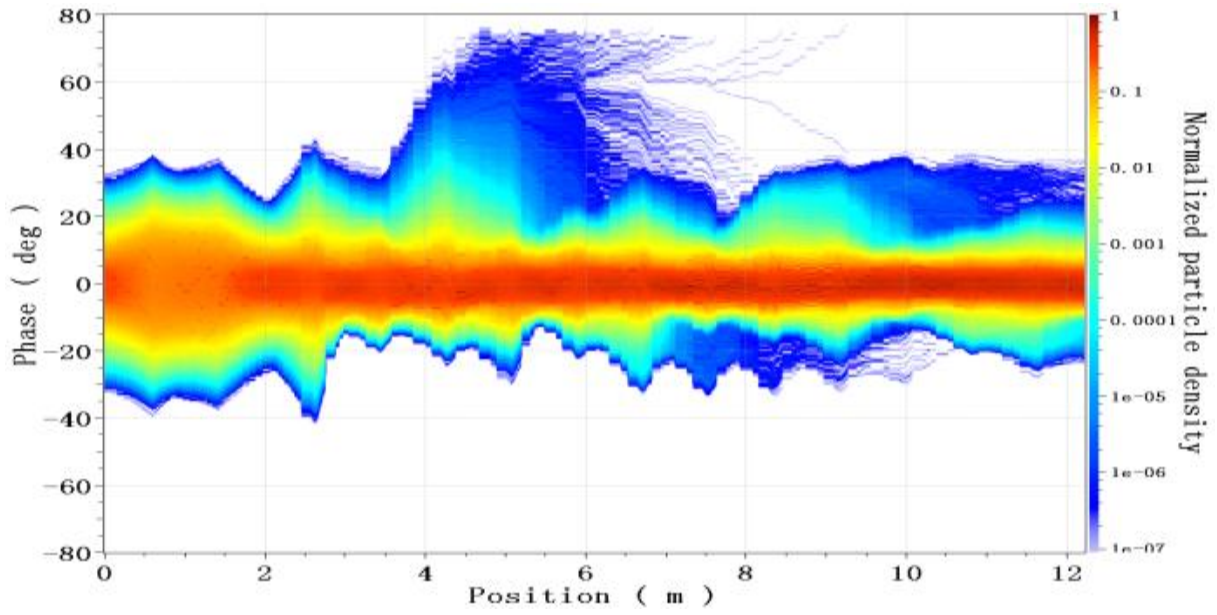

Fig. 8. Particle trajectories in the horizontal and phase planes in the MEBT1 and Spoke012 sections of the injector (The particles exceeding 74° in the phase plane are not shown).

Table 2. Simulation results with different RF errors and with all other errors.

| RF errors sets | | Ex | Ey | Ez | Beam |
|---|---|---|---|---|---|
| Amplitude (%) | Phase (°) | (%) | (%) | (%) | loss rate |
| 0 | 0 | 9.3 | 8.1 | 50 | $7\times10^{-8}$ |
| 0.5 | 0.5 | 9.5 | 8.4 | 51 | $1.2\times10^{-7}$ |
| 1 | 1 | 9.7 | 8.5 | 57 | $1.5\times10^{-7}$ |
| 1.5 | 1.5 | 11.8 | 10.7 | 68 | $2.5\times10^{-7}$ |

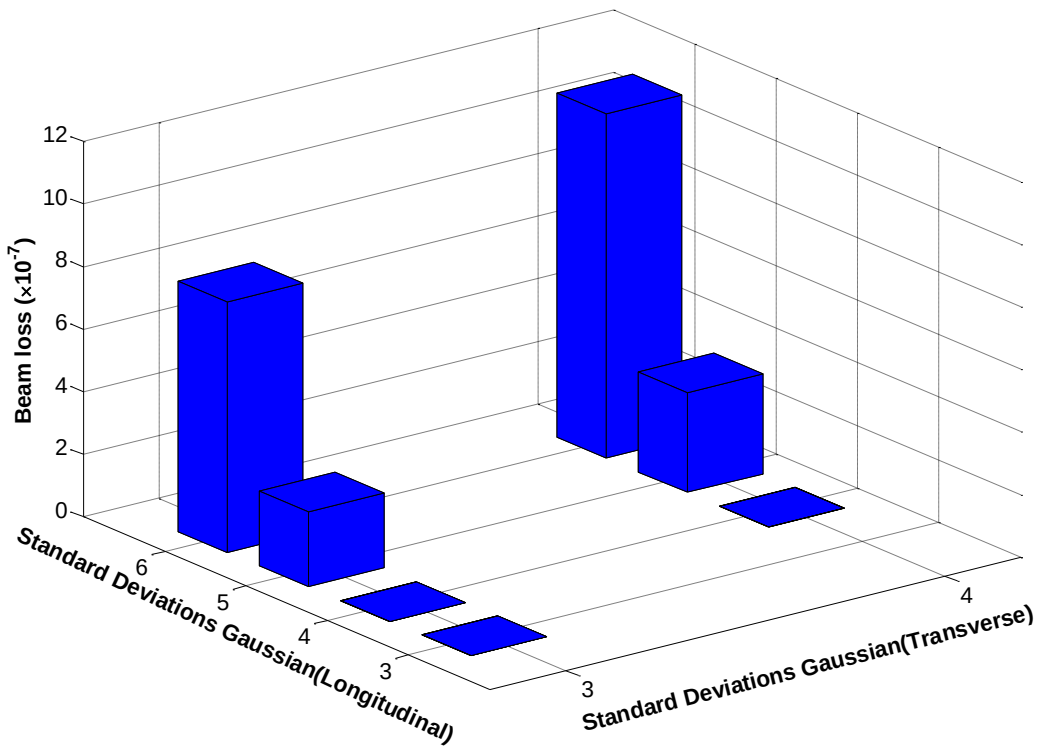

Fig. 9. Beam loss with different initial distributions with errors.

## 3 Improvement of the Spoke012 lattice

As the large total longitudinal emittance from the RFQ is difficult to reduce, it looks that the only solution to minimize beam losses in the Spoke012 section is to enlarge the longitudinal acceptance. Thus an improved lattice with shorter periods has been designed [7], which employs shorter solenoids and smaller synchronous phase (larger in absolute value) with larger acceptance as shown in Fig.10. The solenoid length is decreased from 300 mm to 150 mm and the synchronous phase of the first period is decreased from -37° to -45°. Table 3 shows the comparison results between the nominal and improved designs with errors. The emittance growth is taken from the one at the RFQ exit to the one at the injector exit. The particle trajectories with errors in the horizontal and phase planes are shown in Fig. 11. One can see that with the improved lattice there is no beam loss and the emittance growth is also smaller. This means that it has a better error tolerance.

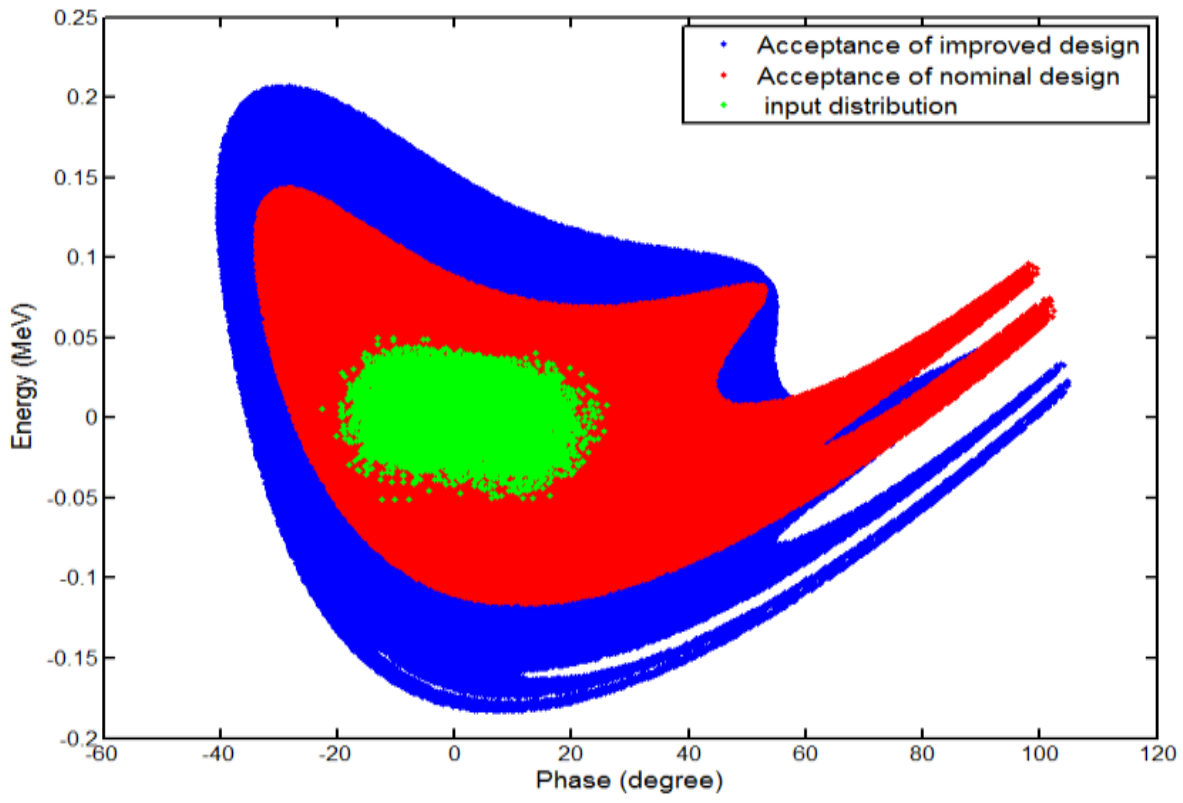

Fig. 10 Comparison between the acceptances of the nominal and the improved lattices.

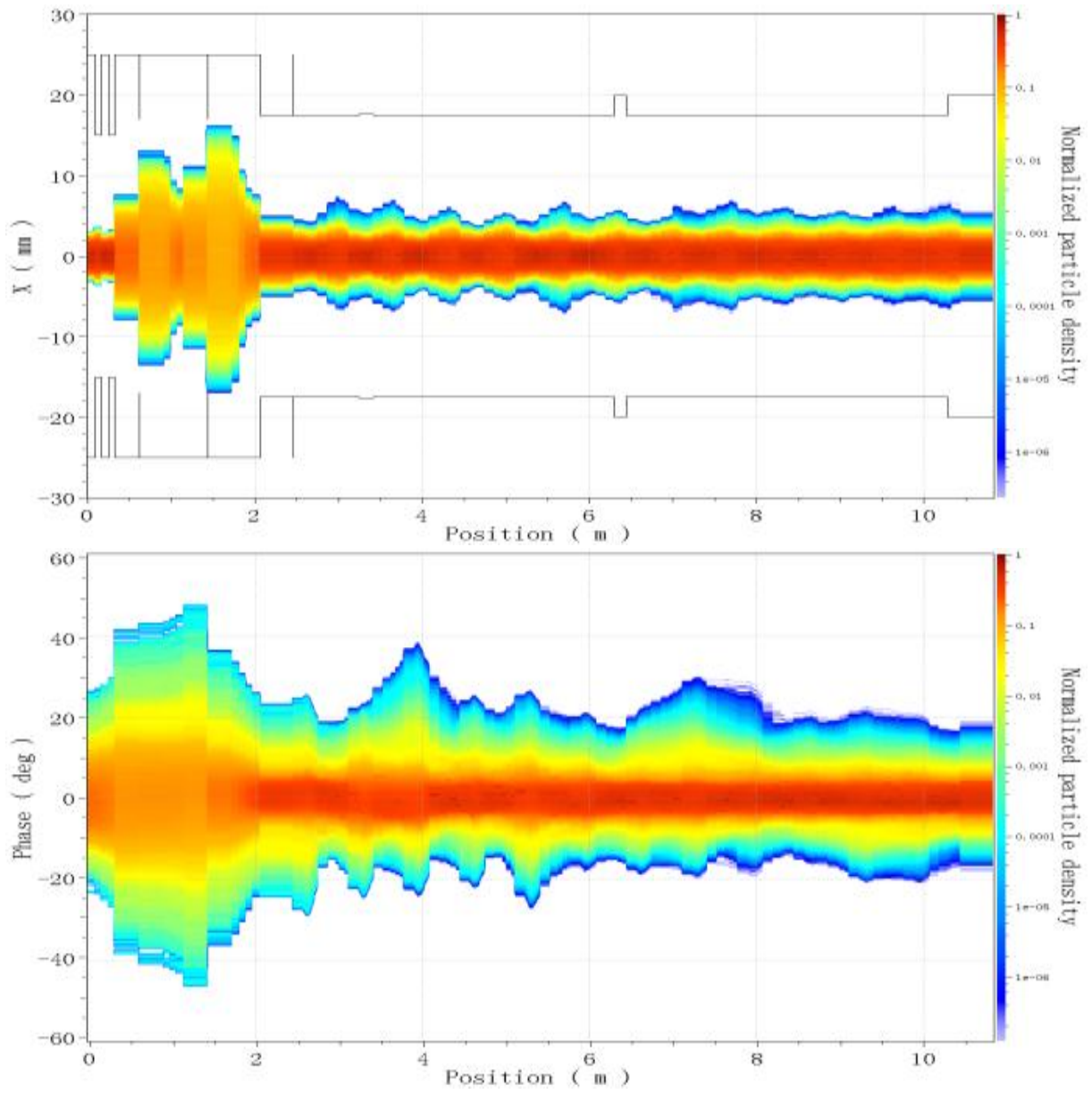

Fig. 11 Particle trajectories in the horizontal plane and phase plane in the injector with an improved Spoke012 lattice.

Table 3: Comparison between the nominal and improved lattices with errors

| Lattice | | Nominal design | Improved design |
|---|---|---|---|
| Ex (%) | Average | 9.5 | 6.1 |
| | rms | 3.7 | 2.3 |
| Ey (%) | Average | 8.9 | 7.7 |
| | rms | 2.5 | 2.4 |
| Ez (%) | Average | 160 | 5.6 |
| | rms | 280 | 2 |
| rms dp/p jitter | | $1.8\times10^{-4}$ | $1.3\times10^{-4}$ |
| rms phase jitter (°) | | 0.4 | 0.3 |
| Beam loss | | $1.7\times10^{-6}$ | 0 |

## 4 Conclusions

With the reasonable error settings to all the elements in the C-ADS injector, the rms residual orbit errors can be controlled within 0.3 mm with orbit correction, but it still has a beam loss rate of $1.7\times10^{-6}$ which is considered mainly from the RF errors and the low longitudinal acceptance. An improved lattice for the Spoke012 section with a shorter period length and a smaller synchronous phase is proven to have better error tolerance. The beam loss rate and emittance growth can be well controlled with the new lattice.

*The authors would like to thank other colleagues in the ADS accelerator physics group for the discussions.*

# C-ADS 注入器 I 的误差分析及优化

孟才, 李智慧, 唐靖宇
中国科学院高能物理研究所，北京，10049，中国

**摘要** C-ADS 注入器方案 I（简称注入器 I）是工作在连续波模式下的流强为 10mA、能量为 10MeV 的质子直线加速器。该注入器主要包括输出能量为 3.2MeV 的常温 4 翼型的 RFQ 和由置于 1 个长恒温器中的 12 个超导轮幅腔组成的超导加速段。本文介绍了注入器的误差对束流动力学的影响，包括准直误差和场误差的误差分析及轨道校正。根据详细的数值模拟和分析给出了合适的校正方案，可以将中心轨道的偏移控制在 0.3mm 以内。模拟计算结果表明，基本设计方案给出的束流损失率为 $1.7\times10^{-6}$ 。在对超导加速段的 Lattice 设计方案进行改进后，获得了了更大的接受度和误差容忍度。